\documentstyle[12pt]{article}
\newcommand{\be}{\begin{equation}}
\newcommand{\ee}{\end{equation}}
\newcommand{\ba}{\begin{eqnarray}}
\newcommand{\ea}{\end{eqnarray}}

\begin{document}
\begin{center}
{\bf DOUBLE  SHAPE  INVARIANCE  OF  TWO-DIMENSIONAL  SINGULAR  MORSE  MODEL}\\
\vspace{0.3cm}

{\large \bf F. Cannata$^{1,}$\footnote{E-mail: cannata@bo.infn.it},
M.V. Ioffe$^{2,}$\footnote{{\it
Corresponding author.}
Phone:+7(812)4284553; FAX: +7(812)4287240; E-mail:
m.ioffe@pobox.spbu.ru}, D.N. Nishnianidze}$^{2,3,}$\footnote{E-mail:
qutaisi@hotmail.com}\\
\vspace{0.2cm}
$^1$ Dipartimento di Fisica and INFN, Via Irnerio 46, 40126 Bologna, Italy.\\
$^2$ Sankt-Petersburg State University, 198504 Sankt-Petersburg, Russia\\
$^3$ Kutaisi Technical University, 4614 Kutaisi, Republic of
Georgia
\end{center}
\vspace{0.2cm}
\hspace*{0.5in}
\begin{minipage}{5.0in}
{\small
A second shape invariance property of the two-dimensional
generalized Morse potential is discovered. Though the potential is not
amenable to conventional separation of variables, the above property allows
to build purely algebraically part of the spectrum
and corresponding wave functions, starting from {\it one}
definite state, which can be obtained by the method of $SUSY$-separation
of variables, proposed recently.\\
}
\end{minipage}
\vspace*{0.2cm}
\section*{\bf 1. \quad Introduction}
\vspace*{0.2cm}
\hspace*{3ex}


Recently the notion of shape invariance \cite{genden} was generalized \cite{two}, \cite{ioffe}
to two-dimensional
cases, which are not amenable to conventional separation of variables in the
Schr\"odinger equation. It was shown that, in contrast to the
one-dimensional situation, in general the shape invariance itself
gives algebraically {\it only part} of spectra (and wave functions), leading to
the partial (quasi-exact) solvability of the model.
The fact that only a partial solution for the spectral problem is provided
by shape invariance for two-dimensional problems is related to the
dependence (in general) of the {\bf ground state energy} on the parameters
of the model (in one-dimensional case one usually had $E_{gs} \equiv 0,$ not depending
on the parameter of shape invariance). Also, possible degeneracy of levels in
two-dimensional models
is important: for complete solvability one has to know both energy levels and {\bf all}
corresponding wave functions.

In particular, the two-dimensional generalized
(singular\footnote{We remark that the behaviour of wave functions at the singularity
is under control (see \cite{two}).})
Morse potential was demonstrated \cite{two}, \cite{ioffe} to
satisfy shape invariance. Furthermore this model allows for
$SUSY$-separation of variables (i.e. separation of variables in the
supercharge), and therefore makes it possible to construct some wave functions as
linear combinations of zero modes of the supercharge. Then each of
these "principal" eigenfunctions was used to generate (purely algebraically!) the shape
invariance chain of excited levels.

In the present paper some additional properties of the two-dimensional
generalized Morse potential are studied (Section 2). It is shown
that besides the shape invariance, described in \cite{two}, the model
possesses an additional shape invariance, which can also be used to
construct corresponding chains of states (Section 3). In this construction
the wave functions of the {\bf first} shape invariance chain of the previous Section play the role
of the "principal" states for new shape invariance chains. Thus it is shown (Section 4)
that the combination of both shape invariances allows to build
{\bf algebraically} the same part of the spectrum
and corresponding eigenfunctions starting from {\bf only one} "principal" state.
The mutual interrelation between the "old" and "new" chains is also
clarified.

\vspace*{0.2cm}
\section*{\bf 2. \quad Shape invariant two-dimensional Morse model}
\vspace*{0.2cm}
\hspace*{3ex}
The direct $d$-dimensional $(d \geq 2)$ generalization of Witten's
SUSY QM includes \cite{abei} Schr\"odinger operators both with
scalar and with matrix potentials. In order to avoid the appearance of matrix
components in the two-dimensional Superhamiltonian, one can explore \cite{david},
\cite{ioffe} for the case $d=2$ the idea of one-dimensional Higher Order
SUSY QM \cite{hsusy} and to take, for example, the second order supercharges
with hyperbolic (Lorentz) metrics:
\be
Q^{\pm} = \partial_1^2 - \partial_2^2 + C_i(\vec x)\partial_i + B(\vec x);
\qquad \vec x = (x_1, x_2); \qquad \partial_i = \partial /\partial x_i;\quad i=1,2.
\label{supercharge}
\ee
The main relations of SUSY QM are the two-dimensional SUSY intertwining relations:
\be
H^{(1)} Q^+ = Q^+ H^{(2)}; \qquad H^{(1,2)} = - \Delta^{(2)} + V^{(1,2)}(\vec x);\quad
\Delta^{(2)}\equiv \partial_i^2,
\label{intertw}
\ee
for which some partial solutions were found \cite{david}
providing nontrivial models with Schr\"odinger Hamiltonians $H^{(1)}, H^{(2)}$,
not allowing separation of variables. It is convenient to consider also light-cone coordinates:
\be
x_{\pm} \equiv x_1 \pm x_2; \qquad \partial_{\pm} = \frac{1}{2}(\partial_1 \pm \partial_2); \quad
C_{\pm} = C_1 \mp C_2,
\nonumber
\ee
where \cite{david} due to (\ref{intertw}) $C_{\pm} = C_{\pm}(x_{\pm}),$ and
\be
\partial_-(C_-F) = - \partial_+(C_+F); \qquad
F(\vec x) \equiv F_1(2x_1) + F_2(2x_2).  \label{CF}
\ee
The potentials $V^{(1,2)}(\vec x)$ are:
\be
V^{(1,2)}(\vec x) = \pm \frac{1}{2} \Bigl(C'_+(x_+) + C'_-(x_-)\Bigr) +
\frac{1}{8} \Bigl(C_+^2(x_+) + C_-^2(x_-)\Bigr) +
\frac{1}{4}\Bigl(F_2(2x_2) - F_1(2x_1)\Bigr), \label{potentials}
\ee
and
\be
B(\vec x) = \frac{1}{4} \Bigl(C_+ C_- + F_1(2x_1) + F_2(2x_2)\Bigr).
\label{BB}
\ee
Recently two specific models of the list of particular solutions of (\ref{CF}) were
shown \cite{two}, \cite{pavel} to be partially (quasi-exactly \cite{turbiner}) solvable,
i.e. part of eigenvalues and eigenfunctions of the system was found
analytically.

The first of the models - the two-dimensional Morse potential - is
characterized\footnote{The constants used
here differ from those in \cite{two},
\cite{ioffe}. The present choice is useful for the derivation below and can be easily made consistent with
the previous ones by suitable shifts of coordinates.} by:
\ba
C_+(x_+)&=&4a\alpha;\qquad C_-(x_-)= 4a\alpha\coth(\frac{\alpha x_-}{2});
\label{CC}\\
\frac{1}{4} F_{1,2}(2x_{1,2}) &=& \mp(2c-\alpha ) \exp{(-\alpha x_{1,2})} \mp
\exp{(-2\alpha x_{1,2})}; \label{F1}
\ea
With this choice the potentials (\ref{potentials}) can be naturally
interpreted as a two-dimensional (non-separable, singular)
generalization of the one-dimensional Morse potential:
\ba
V^{(1,2)}(\vec x; a, \alpha ,c\bigr) &=& \alpha^2a(2a \mp 1)\sinh^{-2}(\alpha x_-/2)
+ (2c - \alpha )\Bigl(\exp(-\alpha x_1)+\exp(-\alpha
x_2)\Bigr)+\nonumber\\
&+&\Bigl(\exp(-2\alpha x_1)+\exp(-2\alpha x_2)\Bigr) + 4a^2\alpha^2.
\label{morse}
\ea

While the Schr\"odinger equations with potential $V^{(1,2)}$ in (\ref{morse})
are not amenable to standard separation of variables, nevertheless, we can apply
the recently proposed \cite{two} (see also \cite{ioffe})
method of $SUSY$-separation of variables (variables are separable not in
$H^{(1,2)},$ but in the supercharge $Q^+$). Then zero modes of $Q^+$
were constructed analytically in terms of hypergeometric functions: suitably chosen
linear combinations of them provide (see details in \cite{two}, \cite{ioffe})
a set of "principal" wave functions $\Psi_{k,0}^{(2)}(\vec x; a,\alpha ,c)$ of the Hamiltonian
$H^{(2)}(\vec x; a,\alpha ,c)$  and
corresponding energy eigenvalues, which depend on arbitrary values of parameters:
\be
E^{(2)}_{k,0}(a,\alpha ,c) =
-2\Bigl[ 2a\alpha^2 s_k - \epsilon_k\Bigr] =
 -2 [-2a\alpha^2(\frac{c}{\alpha}+k)+(c+k\alpha)^2];
\quad k=0,1,2...
\label{energy}
\ee
$\epsilon_k$ and $s_k$ above were defined in \cite{two}, and are now expressed
in terms of new parameter $c$ instead of $A \equiv (c-\frac{\alpha}{2})^2$ (the appropriate
constant shift of $x_{1,2}$ was also used here).

The model of Eq.(\ref{morse}) enjoys \cite{two} an additional remarkable property -
the two-dimensional shape invariance:
\be
H^{(1)}(\vec x; a,\alpha ,c) = H^{(2)}(\vec x; a-\frac{1}{2},\alpha ,c) + {\cal R} (a,\alpha ,c);
\quad {\cal R} (a,\alpha ,c) = \alpha^2(4a-1).
\label{shape}
\ee
Similarly to the one-dimensional shape invariance, each "principal" eigenfunction
$\Psi^{(2)}_{k,0}$
gives start to a whole shape invariance chain of eigenstates of $H^{(2)}(\vec x; a,\alpha ,c),$
which can be built by means of a sequence of supercharges $Q^-\equiv (Q^+)^{\dagger}:$
\ba
\Psi_{k,m}^{(2)}(\vec x; a,\alpha ,c) &=& Q^-(a,\alpha ,c)\cdot
Q^-(a-\frac{1}{2},\alpha ,c)\cdot
Q^-(a-1,\alpha ,c)\cdot ...\nonumber\\&... & \cdot
Q^-(a-\frac{1}{2}(m-1),\alpha ,c) \Psi_{k,0}^{(2)}(\vec x; a-\frac{1}{2}m,\alpha ,c);\quad
(m=1,2,...),
\label{states}
\ea
with eigenvalues:
\ba
E^{(2)}_{k,m}(a,\alpha ,c) &=& E^{(2)}_{k,0}(a-\frac{1}{2}m,\alpha ,c) +
{\cal R} (a-\frac{1}{2}(m-1),\alpha ,c) +
... + {\cal R} (a,\alpha ,c)=\nonumber\\
&=& 2(c+k\alpha )\Bigl(2a\alpha - a\alpha m - (c+k\alpha )\Bigr) + \alpha^2m(4a-m).
\label{energies}
\ea
The sequence in (\ref{states}) is constrained only by the condition of normalizability
of these wave functions.

\vspace*{0.2cm}
\section*{\bf 3. \quad The new shape invariance of the Morse potential}
\vspace*{0.2cm}
\hspace*{3ex}
In order to demonstrate the existence of a second invariance of the two-dimensional
Morse potential (\ref{morse}), we link it to another two-dimensional
system, found in \cite{david} among solutions (\ref{CF}) - (\ref{BB}) of intertwining relations
with hyperbolic (Lorentz) metrics in supercharges.
The corresponding functions will be denoted by tildes and
their arguments by $\vec y=(y_1, y_2)$:
\ba
\widetilde{C}_+(y_+)=4\Bigl(\exp(\alpha y_+)+c\Bigr);\quad
\widetilde{C}_-(y_-)=4\Bigl(\exp(\alpha y_-)+c\Bigr); \quad
y_{\pm}\equiv y_1\pm y_2\nonumber\\
\widetilde{F}_1(2y_1)=0; \quad \frac{1}{4}\widetilde{F}_2(2y_2)=2d\Bigl(\exp(\alpha
y_2)-\exp(-\alpha y_2)\Bigr)^{-2};\nonumber\\
\widetilde B=4\Bigl(\exp(\alpha y_-)+c\Bigr)\cdot\Bigl(\exp(\alpha y_+)+c\Bigr) +
2d\sinh^{-2}(\alpha y_2).\nonumber
\ea
The superpartner potentials found in \cite{david}:
\ba
\widetilde{V}^{(1,2)}(\vec y; a,\alpha ,c) &=&
2\Bigl(\exp(2\alpha y_+)+\exp(2\alpha y_-)\Bigr)+
2(2c\pm \alpha)\Bigl(\exp(\alpha y_+)+\exp(\alpha y_-)\Bigr)\nonumber\\
 &+&2d\cdot\sinh^{-2}(\alpha y_2),\label{6}
\ea
satisfy to the intertwining relations:
\ba
\widetilde{H}^{(1)}(\vec y; a,\alpha ,c)\widetilde{Q}^+(\vec y; a,\alpha ,c)=
\widetilde{Q}^+(\vec y; a,\alpha ,c)\widetilde{H}^{(2)}(\vec y; a,\alpha ,c);\nonumber\\
\widetilde{Q}^+(\vec y; a,\alpha ,c)=
\partial^2_{y_1}-\partial^2_{y_2}+\widetilde{C}_+(y_+)\partial_{y_-}+
\widetilde{C}_-(y_-)\partial_{y_+}+\widetilde B.\nonumber
\ea

In order to relate this system to the model of Section 2, one has to replace:
\ba
y_{+} &\equiv & -x_1; \quad y_- \equiv -x_2; \qquad d
\equiv \alpha^2 a(2a+1); \nonumber\\
\widetilde{H}^{(1,2)}(\vec y(\vec x); a,\alpha ,c) &\equiv & 2 h^{(1,2)}(\vec x; a,\alpha ,c);
\quad \widetilde Q^+(\vec y(\vec x); a,\alpha ,c)\equiv q^+(\vec x; a,\alpha ,c)
\nonumber
\ea
The components $h^{(1,2)}(\vec x; a,\alpha ,c)$ of new Superhamiltonian become:
\ba
h^{(1,2)}(\vec x; a,\alpha ,c)&=& -\Delta^{(2)} + \alpha^2a(2a+1)\sinh^{-2}(\alpha x_-/2)
+ 4a^2\alpha^2 +\nonumber\\
&+&(2c \pm \alpha)\Bigl(\exp(-\alpha
x_1)+\exp(-\alpha x_2)\Bigr)+
\Bigl(\exp(-2\alpha x_1)+\exp(-2\alpha x_2)\Bigr),
\label{new}
\ea
and they are intertwined:
\be
h^{(1)}(\vec x; a,\alpha ,c) q^+(\vec x; a,\alpha ,c) =
q^+(\vec x; a,\alpha ,c) h^{(2)}(\vec x; a,\alpha ,c)
\label{13}
\ee
 by the supercharge:
\ba
q^+(\vec x; a,\alpha ,c)&=&4\partial_1\partial_2 - 4\Bigl(\exp(-\alpha x_1) + c\Bigr)\partial_2
-4\Bigl(\exp(-\alpha x_2) + c\Bigr)\partial_1 +\nonumber\\ &+&
4\Bigl(\exp(-\alpha x_1) + c\Bigr)\cdot\Bigl(\exp(-\alpha x_2) + c\Bigr)
+ 2\alpha^2 a(2a+1)\sinh^{-2}(\alpha x_-/2). \label{11}
\ea
From (\ref{new}) one can conclude that this supersymmetrical
system has also the shape invariance property, but in this case in the parameter $c$:
\be
v^{(1)}(\vec x; a,\alpha ,c) = v^{(2)}(\vec x; a,\alpha ,c + \alpha)
\label{17}
\ee
where the term, analogous to $\cal{R}$ in (\ref{shape}), now vanishes.
Therefore also the spectrum of $h^{(2)}(\vec x; a,\alpha ,c)$ can be
obtained algebraically:
starting from some "principal" (for this model) wave function with
energy $e^{(2)}_{l,0}(a, \alpha , c)\quad l=0,1,2,..,$
one will obtain the shape invariance chain of states with energies
\be
e^{(2)}_{l,n}(a, \alpha , c) = e^{(2)}_{l,0}(a, \alpha , c+n\alpha)\qquad  n=1,2,... .
\label{newE}
\ee
As for the choice of "principal" states for this model, it is convenient to choose states
of the {\bf first} shape invariance chain of Section 2:
$e^{(2)}_{k,0}(a, \alpha , c) = E^{(2)}_{0,k}(a,\alpha ,c).$

Comparing the two supersystems (\ref{morse}) and (\ref{new}), one notices that:
\be
H^{(2)}(\vec x; a,\alpha ,c) = h^{(2)}(\vec x; a,\alpha ,c),
\label{coincide}
\ee
therefore the same two-dimensional system (\ref{morse}) - with generalized Morse potential -
participates in two different intertwining relations (\ref{intertw}) and (\ref{13}) and
possesses also two independent shape invariance properties (\ref{shape}) and (\ref{17}),
expressed in transformations of parameters $a$ and $c,$ respectively.

\vspace*{0.2cm}
\section*{\bf 4. \quad The spectrum of the singular Morse potential}
\vspace*{0.2cm}
\hspace*{3ex}
A priori, one has to expect, that each shape invariance will give rise to
its own
chain of states and corresponding energies. But a more careful analysis shows that
these chains include states which are overlapping. First of all, one can observe this
overlap from the explicit formulas
(\ref{energies}) and (\ref{newE}) for the spectra. But due to the
possible (in principle) degeneracy of levels in
two-dimensional Quantum Mechanics, it still does not imply the coincidence of
wave functions. Nevertheless, this coincidence holds as can
be straightforwardly checked by means of the operator equality:
\be
q^-(a,\alpha ,c)\cdot Q^-(a,\alpha ,c+\alpha) = Q^-(a,\alpha ,c)\cdot
q^-(a-\frac{1}{2},\alpha ,c); \quad
q^-(\vec x; a,\alpha ,c)
\equiv (q^+(\vec x; a,\alpha ,c))^{\dagger}.
\label{lattice}
\ee
For example, just the operators in both sides of this equation, acting onto the wave function
$\Psi^{(2)}_{0,0}(a-\frac{1}{2}, \alpha , c+\alpha),$ give {\it by two different ways}
the eigenfunction $\Psi^{(2)}_{1,1}(a, \alpha , c)$ with energy $E^{(2)}_{1,1}(a, \alpha , c).$
Analogously, one can check that the wave function $\Psi^{(2)}_{k,m}$ with arbitrary pair
of indices $(k,m)$
can be obtained by different ways, via chains of operators $Q^-$ and $q^-,$ giving the same
result due to equalities similar to (\ref{lattice}).

A better understanding of the above mentioned overlap can be obtained by realizing that the
"principal" states of the new shape invariance (i.e. the states from which one starts
the construction of chains) are chosen to be the
states $E^{(2)}_{0,m}(a,\alpha ,c)=e^{(2)}_{m,0}(a,\alpha ,c)$
(see (\ref{energies})). So, the second shape invariance
acts transversely in respect to the first. Acting with $q^-(\vec x; a,\alpha ,c)$
(see (\ref{11})) $k$ times on a generic state
$\Psi^{(2)}_{0,m}(a,\alpha ,c)$ leads to the state $\Psi^{(2)}_{k,m}.$
Thus the overlap can be concisely depicted as
\be
E^{(2)}_{k,m}=e^{(2)}_{m,k}.\nonumber
\ee
In other words, one can move first "up" $c \rightarrow c + \alpha $ and then "transversely"
$ a \rightarrow a-\frac{1}{2},$ or first transversely and then up on the "energy lattice".

\vspace*{0.2cm}
\section*{\bf 5. \quad Remarks and conclusions}
\vspace*{0.2cm}
\hspace*{3ex}
We stress that each Hamiltonian, participating in SUSY intertwining relations
of second order (\ref{intertw}), is integrable \cite{david}, \cite{ioffe},
i.e. it has a symmetry operator (integral of motion) of fourth order in derivatives:
\be
R^{(1)} = Q^+Q^-;\qquad R^{(2)} = Q^-Q^+; \qquad [H^{(i)}, R^{(i)}]=0; \quad i=1,2,
\nonumber
\ee
which {\bf cannot} be reduced by elimination of operator functions of the Hamiltonian.
It can be checked straightforwardly, though rather tediously, that the symmetry operators
associated to intertwining relations (\ref{13}) {\it do not} give a new, {\it independent},
symmetry operator for the Hamiltonian $H^{(2)},$ indeed:
\be
q^-q^+ + Q^-Q^+ = 4\Bigl(H^{(2)}+2c^2\Bigr)^2 + 16c^2(c^2-4a^2\alpha^2).
\label{inter}
\ee

The gluing condition (\ref{coincide}) of two $SUSY$ systems (\ref{morse}) and (\ref{new})
provides an opportunity to link the components $H^{(1)}$ and $h^{(1)}$ by intertwining
operators of fourth order in derivatives:
\be
H^{(1)} \Bigl(Q^+\cdot q^-\Bigr) = \Bigl(Q^+\cdot q^-\Bigr) h^{(1)}.\nonumber
\ee
This intertwining produces also a shape invariance involving two parameters :
\be
H^{(1)}(\vec x; a,\alpha ,c) = h^{(1)}(\vec x; a-\frac{1}{2},\alpha ,c-\alpha )
+ {\cal R}(a,\alpha, c).
\nonumber
\ee
The associated symmetry operators for $H^{(1)}$ of order eight
$\Bigl(Q^+\cdot q^-\Bigr)\cdot \Bigl(q^+\cdot Q^-\Bigr)$
{\it are reducible} to a polynomial function of $H^{(1)}$ and $R^{(1)}.$

Let us note that treatments, similar to the ones of Sections 2 - 4, can be applied
also to the complexified version of singular
two-dimensional Morse potential (\ref{morse}) (see \cite{pseudo}), including a
complexified version of the related two-dimensional model (\ref{6}) and leading to
a complex form of double shape invariance.

In conclusion, the main results of the paper are the following.

- A new property of two-dimensional Morse model - the second shape
invariance - was found and investigated.

- The excited states of new shape invariance chains were proved to
coincide with the states obtained from the first shape invariance.

- It was shown that the same set of states of partially solvable
two-dimensional Morse model can be built now from {\bf only one}
"principal" state $\Psi^{(2)}_{0,0}(a,\alpha ,c)$, i.e. one needs {\bf only the first}
zero mode of the supercharge $Q^+$ to be obtained by the method of SUSY-separation of
variables. This is much easier to build.

- Though the Morse Hamiltonian (11) obeys two different intertwining
relations and has, correspondingly, two fourth order symmetry operators,
the system {\bf is not superintegrable}, since these operators are inter-related
by the hamiltonian (see (\ref{inter})).

\section*{\bf Acknowledgements}
\vspace*{0.2cm}
\hspace*{3ex}
M.V.I. and D.N.N. are grateful to the University of Bologna and INFN for
support and kind hospitality. This work was partially supported by the Russian
Foundation for Fundamental Research (Grant No.05-01-01090).


\vspace{.2cm}

\begin{thebibliography}{}
\bibitem{genden} L.E.Gendenshtein 1983 {\it JETP Lett.} {\bf 38} 356
\bibitem{two} F. Cannata, M.V. Ioffe, D.N. Nishnianidze 2002
{\it J.Phys.:Math.Gen.} {\bf A35} 1389
\bibitem{ioffe} M.V. Ioffe 2004 {\it J.Phys.:Math.Gen.} {\bf A37} 10363
\bibitem{abei}
A.A. Andrianov, N.V. Borisov, M.V. Ioffe 1984 {\it JETP Lett.}
{\bf 39} 93\\
A.A. Andrianov, N.V. Borisov, M.V. Ioffe 1984 {\it Phys. Lett.}
{\bf A105} 19\\
A.A. Andrianov, N.V. Borisov, M.V. Ioffe 1985 {\it Theor. Math.Phys.}
{\bf 61} 1078\\
A.A. Andrianov, N.V. Borisov, M.V. Ioffe, M.I. Eides 1985
{\it Phys. Lett.} {\bf 109A} 143\\
A.A. Andrianov, N.V. Borisov, M.V. Ioffe, M.I. Eides 1984
{\it Theor. Math. Phys.} {\bf 61} 965
\bibitem{david}  A. Andrianov, M. Ioffe, D. Nishnianidze 1995
{\it Phys.Lett.}, {\bf A201} 103\\
A.A. Andrianov, M.V. Ioffe, D.N. Nishnianidze 1995 {\it Theor.Math.Phys.}
{\bf 104} 1129\\
A.A. Andrianov, M.V. Ioffe, D.N. Nishnianidze 1996 solv-int/9605007;
Published in: 1995 {\it Zapiski Nauch.
Seminarov POMI RAN} ed.L.Faddeev et.al. {\bf 224} 68 (In Russian);\\
A.A. Andrianov, M.V. Ioffe, D.N. Nishnianidze 1999
{\it J.Phys.:Math.Gen.} {\bf A32} 4641
\bibitem{hsusy} A.A. Andrianov, M.V Ioffe, V.P. Spiridonov 1993 {\it Phys.Lett.}
{\bf A174} 273\\
A.A. Andrianov, F. Cannata, J.-P. Dedonder, M.V. Ioffe 1995
{\it Int.J.Mod.Phys.} {\bf A10} 2683\\
B.F. Samsonov 1996 {\it Mod. Phys. Lett.} {\bf A11} 1563\\
D.J. Fernandez C., M.L. Glasser, L.M. Nieto 1998 {\it Phys.Lett.} {\bf A240} 15\\
V.G. Bagrov, B.F. Samsonov, L.A. Shekoyan 1998 quant-ph/9804032\\
S. Klishevich, M. Plyushchay 1999 {\it Mod. Phys. Lett.} {\bf A14} 2739\\
A.A. Andrianov, F. Cannata, M.V. Ioffe, D.N.Nishnianidze 2000
{\it Phys.Lett.} {\bf A266} 341\\
D.J. Fernandez C., J. Negro, L.M. Nieto 2000 {\it Phys.Lett.} {\bf A275} 338\\
S. Klishevich, M. Plyushchay 2001 {\it Nucl.Phys.} {\bf B606[PM]} 583\\
H. Aoyama, M. Sato, T. Tanaka 2001 {\it Phys.Lett.} {\bf B503} 423\\
H. Aoyama, M. Sato, T. Tanaka 2001 {\it Nucl.Phys.} {\bf B619} 105\\
R. Sasaki, K. Takasaki 2001 {\it J. Phys. A: Math.Gen.} {\bf 34} 9533\\
A.A. Andrianov, A.V. Sokolov 2003 {\it Nucl.Phys.} {\bf B660} 25\\
A.A. Andrianov, F. Cannata 2004 {\it J. Phys. A: Math.Gen.} {\bf 37} 10297\\
M.V. Ioffe, D.N. Nishnianidze 2004 {\it Phys.Lett.} {\bf A327} 425
\bibitem{pavel} M.V. Ioffe, P.A. Valinevich 2005 {\it J. Phys. A: Math.Gen.} {\bf 38} 2497
\bibitem{turbiner} A.V. Turbiner 1988 {\it Comm.Math.Phys.} {\bf 118} 467\\
A.G.Ushveridze 1989 {\it Sov.J.Part.Nucl.} {\bf 20} 504
\bibitem{pseudo} F. Cannata, M.V. Ioffe, D.N. Nishnianidze 2003
{\it Phys.Lett.} {\bf A310} 344

\end{thebibliography}
\end{document}